\documentclass[twocolumn,english,showpacs,preprintnumbers,amsmath,amssymb,floatfix]{revtex4}
\usepackage[T1]{fontenc}
\usepackage[latin9]{inputenc}
\usepackage{color}
\usepackage{array}
\usepackage{amstext}
\usepackage{graphicx}
\usepackage{esint}

\makeatletter

\@ifundefined{textcolor}{}
{%
 \definecolor{BLACK}{gray}{0}
 \definecolor{WHITE}{gray}{1}
 \definecolor{RED}{rgb}{1,0,0}
 \definecolor{GREEN}{rgb}{0,1,0}
 \definecolor{BLUE}{rgb}{0,0,1}
 \definecolor{CYAN}{cmyk}{1,0,0,0}
 \definecolor{MAGENTA}{cmyk}{0,1,0,0}
 \definecolor{YELLOW}{cmyk}{0,0,1,0}
 }

\@ifundefined{definecolor}
 {\usepackage{color}}{}
\@ifundefined{definecolor}
 {\usepackage{color}}{}
\makeatother

\makeatother

\usepackage{babel}

\begin{document}

\title{Fragmentation functions of $g\rightarrow \eta_c (^{1}S_0)$ and $g\rightarrow J/\psi (^{3}S_1)$ 
considering the role of heavy quarkonium spin}

\author{S. M. Moosavi Nejad$^{a,b}$}
\email{mmoosavi@yazd.ac.ir}

\affiliation{$^{(a)}$Faculty of Physics, Yazd University, P.O. Box
89195-741, Yazd, Iran}

\affiliation{$^{(b)}$School of Particles and Accelerators,
Institute for Research in Fundamental Sciences (IPM), P.O.Box
19395-5531, Tehran, Iran}

\date{\today}

\begin{abstract}

The production of heavy quarkonia is a powerful tool to test our understanding
of strong interaction dynamics. It is well-known that
the dominant production mechanism for heavy quarkonia with large transverse momentum is fragmentation.
In this work we, analytically, calculate the QCD leading order contribution to 
the process-independent fragmentation functions (FFs) for a gluon to split 
into the vector ($J/\psi$) and pseudoscalar ($\eta_c$) $S$-wave 
charmonium states.
The analyses of this paper differ in which we present, for the first time, an analytical form of the
$g\rightarrow J/\psi$ FF using a different approach (Suzuki's model) in comparison
with other results presented in literatures, where the Braaten's  scheme was used and 
the two-dimensional integrals were presented for the gluon FFs which must be evaluated numerically.
The universal fragmentation probability for the $g\rightarrow J/\psi$ is about $10^{-6}$ which is in good consistency
with the result obtained in the  Braaten's model.


\end{abstract}

\pacs{13.87.Fh; 14.70.Dj; 12.39.Jh; 13.90.+i;  13.60.Le; 12.38.Bx}

\maketitle

\section{Introduction}
\label{sec:intro}
 
Heavy quarkonia, as the bound states of a heavy quark and antiquark, are the simplest 
particles when the strong interactions are concerned.
Their production  has a long history of theoretical calculations and experimental measurements \cite{Brambilla:2010cs}, especially, 
their production at high-energy colliders has been the subject of considerable interest during the past few
years. Nevertheless, the production of heavy quarkonium is still puzzling us after almost forty years since the discovery of $J/\psi$. 
New data have been taken at $e^+e^-$, $ep$ and $p\bar{p}$ colliders, and a wealth of fixed-target data also exist
and with the advances in theory and tremendous amount of precise data
from the Large Hadron Collider (LHC), it is an excellent time to study the physics of heavy quarkonium production.\\
Nowadays, it is well-known that the dominant mechanism to produce the heavy quarkonia at  high
transverse momentum is fragmentation; the production of a parton with a large transverse momentum
which subsequently decays to form a jet containing the expected hadron \cite{Braaten:1993rw}.
It is hence important to obtain the corresponding fragmentation function (FF), in order to properly 
estimate the production rate of a specific quarkonium state.
Because of the simple internal structure of heavy quarkonia 
the perturbative QCD approximations to their FFs are well-defined in the 
nonrelativistic QCD (NRQCD) factorization framework \cite{Bodwin:1994jh}.
To calculate the FFs, beside the current phenomenological approaches 
which are based on the $\chi^2$ analysis of experimental data (see our previous work \cite{Soleymaninia:2013cxa}), 
there are two theoretical schemes which are based on the fact that the FFs for hadrons containing heavy quarks
can be computed analytically using the perturbative QCD (pQCD) \cite{Chang:1991bp,Braaten:1993mp}.
In these theoretical schemes the QCD improved parton model provides a great theoretical frame to extract the FFs.\\
The first scheme for the inclusive production of heavy quarkonium
has been developed by Bodwin, Braaten, and Lepage, who 
proposed a method on the basis of pQCD and the nonrelativistic quark 
model \cite{Bodwin:1994jh,Chang:1991bp,Braaten:1993mp,Braaten:1994bz}.
In this model, the FF of a heavy quark $Q$ into the pseudoscalar or vector heavy-light mesons $Q\bar{q}$  
is defined as the cross section for producing a $Q\bar{q}$-meson plus a light quark $q$ with total 
four-momentum $K^\mu$, divided by the cross section for producing an on-shell $Q$ with the same 
three-momentum $\bold{K}$, while $K_0\rightarrow \infty$.
The fragmentation function in the Braaten's model is defined as \cite{Braaten:1994bz}
\begin{eqnarray}\label{Braaten}
D(z, \mu_0)&=&\frac{1}{16\pi^2}\int ds \quad lim_{K_0\rightarrow \infty}\frac{\sum|\textit{M}|^2}{\sum|\textit{M}_0|^2},
\end{eqnarray}
where $s=K^2$ is the invariant mass of the meson,
$\textit{M}$ is the matrix element for producing mesons plus a light quark $q$, and
$\textit{M}_0$ is the matrix element for producing an on-shell $Q$.
Here, $D$ stands for the fragmentation function and $z$ is the fragmentation 
parameter which refers  to the longitudinal momentum fraction of the quarkonium 
and $\mu_0$ is a fragmentation scale.
This model was applied to the fragmentation processes $\bar{b}\rightarrow B_c$ and
$\bar{b}\rightarrow B^{\star}_c$ in Ref.~\cite{Braaten:1993jn}.\\
Another elaborate model is proposed by Suzuki \cite{Suzuki:1977km,Suzuki:1985up} which is  
based on the convenient Feynman diagrams and the wave function of the respective heavy meson,
where the wave function includes the effect of long-distance in the fragmentation.
In this model, the heavy FFs are calculated using a diagram similar to that in Fig.~\ref{feyn2},
so the analytical expression of FFs depends on the transverse momentum $p_T$ of the parton which 
appears as a phenomenological parameter (e.g. see Eq.~(\ref{d1})), while 
in the Braaten's model the integrations over all freedom degrees are performed.\\
Note that, as soon as more than one hadron is appeared in a hard scattering process, it is 
necessary to take into account the transverse momentum $p_T$ of the partons.
For example, transverse momentum dependent FFs
show up explicitly in several semi-inclusive cross sections, in particular in azimuthal
asymmetries.  In the calculation of QCD corrections to these cross sections, the inclusion of the $p_T$
dependent FFs will be essential. 
Experimentally, based on the 1992-1993 run (run 1A), the CDF collaboration published data on their first measurement of the B-meson
differential cross section $d\sigma/dp_T$ for the exclusive decays $B^+\rightarrow J/\psi K^+$ and
$B^0\rightarrow J/\psi K^{*0}$ \cite{Abe:1995dv}. In \cite{Binnewies:1998vm}, the $p_T$ distribution $d\sigma/dp_T$ is 
considered and the theoretical predictions are compared with the CDF data \cite{Abe:1995dv} for which 5GeV$<p_T<$20GeV.
This is our main motivation to study the transverse momentum dependent FFs in the Suzuki's model.
In Ref.~\cite{Nejad:2013vsa}, using this model we computed an exact
analytical expression of the fragmentation function for c-quark to split into the $D^+/D^0$ mesons to LO.
There, we investigated that there is an excellent consistency between our result and the current well-known phenomenological
models and  also with the experimental data form BELLE and CLEO.\\
In high energy processes, the main contribution of charmonium production results from gluon 
fragmentation, while the charm quark fragmentation contribution is
much too small \cite{Roy:1994ie,Falk:1993rj}. This point was confirmed by the comparison between the 
theoretical predictions and the CDF $J/\psi$ production data. 
Therefore, in this work using the Suzuki's model we focus on the gluon fragmentation into a vector ($J/\psi$) and 
pseudoscalar ($\eta_c$) S-wave charmonium  to leading order of perturbative QCD
and we present, for the first time, an analytical expression
for the transverse momentum dependent fragmentation functions of $g\rightarrow \eta_c, J/\psi$.\\
Note that, in the past few years  the $g\rightarrow \eta_c, J/\psi$ FFs
have been calculated numerically, using the Braaten's model \cite{Braaten:1993rw,Qi:2007sf}.
In these papers, due to the lengthy and cumbersome expression of the $g\rightarrow J/\psi$ FF
the two-dimensional integrals have been presented that must be evaluated numerically.
Moreover, since in the Braaten's model integrations over all freedom degrees are performed, 
then the presented FFs are independent of the transverse momentum of the initial gluon.\\
Our analytical expression for the $g\rightarrow J/\psi$ FF 
will be compared with the numerical result presented in \cite{Qi:2007sf}.
Using the FF obtained, we also calculate the first two moments of FF which
are of phenomenological interest and subject to experimental determination.
They correspond to the $g\rightarrow J/\psi$ branching fraction
and the average energy fraction of the $J/\psi$ meson which receives from the gluon.\\
This paper is organized as follows.
In Sec.~\ref{sec:two}, we explain
our theoretical approach to calculate the FFs using the pQCD.
Our analytical results of the $g\rightarrow \eta_c, J/\psi$ FFs will be 
presented in the Suzuki's model. 
In Sec.~\ref{sec:three}, our numerical results for the gluon FFs are presented. 
Our conclusion is summarized in Sec.~\ref{sec:four}.

\section{Gluon fragmentation into S-wave charmonium: Perturbative QCD scheme}
\label{sec:two}

\begin{figure}
\begin{center}
\includegraphics[width=0.50\linewidth,bb=199 530 392 715]{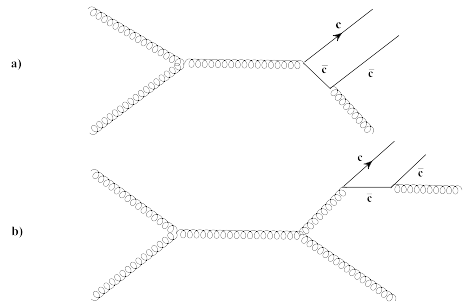}
\caption{\label{feynmangraph}%
Feynman diagrams that contribute to charmonium production: (a) $gg\rightarrow c\bar{c}g$ at order-$\alpha_s^3$, 
(b) $gg\rightarrow c\bar{c}gg$ at order-$\alpha_s^4$ \cite{Braaten:1993rw}.}
\end{center}
\end{figure}
\begin{figure}
\begin{center}
\includegraphics[width=0.50\linewidth,bb=199 535 392 700]{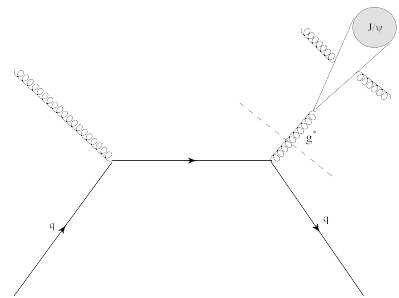}
\caption{\label{feynman}%
Feynman diagram of process $qg\rightarrow qg^\star \rightarrow J/\psi gg$ \cite{Qi:2007sf}.}
\end{center}
\end{figure}
The theoretical schemes for calculating the charmonium FFs are based on the fact that 
the FFs for hadrons containing a heavy quark can be computed theoretically using
perturbative QCD \cite{Braaten:1993rw,Chang:1991bp,Braaten:1993mp,Ma:1997yq}.
The $g\rightarrow \eta_c, J/\psi$ FFs have been already calculated in Refs.~\cite{Braaten:1993rw,Qi:2007sf}, 
using the Braaten's model where the fragmentation function is defined as in (\ref{Braaten}). 
In Ref.~\cite{Braaten:1993rw}, authors considered the typical Feynman diagrams contributed to the 
production of charmonium states at the order $\alpha_s^3$ (Fig.~\ref{feynmangraph}a) and at the 
order $\alpha_s^4$ (Fig.~\ref{feynmangraph}b). They have pointed out that in most regions 
of phase space, the virtual gluons in Fig.~\ref{feynmangraph}b are off their mass shells by amounts of order $p_T$,
where $p_T$ refers to the large transverse momentum of charmonium state, and 
the contribution from this diagram is suppressed relative to the diagram in Fig.~\ref{feynmangraph}a by
a power of $\alpha_s(p_T)$ when the spin-singlet S-wave charmonium  $\eta_c (^{1}S_0)$ is considered. 
Then, they have obtained the $g\rightarrow \eta_c$ FF at $\alpha_s(\mu=2m_c)^2$-order. 
Authors have also calculated the FF for a gluon into $J/\psi$ 
considering the Feynman diagram for the process $g^\star \rightarrow J/\psi gg$ (Fig.~\ref{feynmangraph}b)
to $\alpha_s^3$. They have not presented 
an analytical form for the $D_g^{J/\psi}(z,\mu)$ and, instead,
the result is given in a  two-dimensional integral form that must
be evaluated numerically. In \cite{Qi:2007sf}, using the Braaten's model 
authors have also obtained an integral form for the polarized and unpolarized initial FFs of
gluon into the $J/\psi$ at $\alpha_s(\mu=2m_c)^3$, considering a specific physical process 
shown in Fig.~\ref{feynman}.\\
Here, we focus on the gluon fragmentation into a
point-like $c\bar c$ pair in the $^{1}S_0(\eta_c)$ and $^{3}S_1(J/\psi)$ states
and derive, for the first time, an analytical form of 
their transverse momentum dependent FFs,
using the Suzuki's model.
The leading contribution to the short-distance scattering amplitude of the fragmentation process $g\rightarrow J/\psi$
arises from the partonic process $g\rightarrow c\bar{c}gg$ (Fig.~\ref{feyn2}) and is of order $\alpha_s^3$,
as its dominant decay process is $J/\psi\rightarrow 3g$.
The hard scattering amplitude of the fragmentation  $g\rightarrow \eta_c$ 
results from the process $g\rightarrow c\bar{c}g$ (Fig.~\ref{feyn1}) and is of order $\alpha_s^2$.\\
The fragmentation  parameter $z$ is normally defined in  the Lorentz boost invariant form as $z=(E^H+p_L^H)/(E^g+p_L^g)$,
which is known as the light-cone form.
This definition is hard to  be employed in the application of the gluon FF, because it involves the
transverse momentum of the resulting heavy quarkonium.
Instead, the following non-covariant definition is usually used approximately
\begin{eqnarray}\label{parameter}
z=\frac{E^H}{E^g}\cdot
\end{eqnarray}
When the fragmenting gluon momentum $|\vec{k}|\rightarrow \infty$, the definition (\ref{parameter})
is equivalent to the light-cone definition \cite{Qi:2007sf}. 
Therefore, we adopt the infinite momentum frame where
the fragmentation  parameter is defined as in (\ref{parameter}).
In Ref.~\cite{Qi:2007sf}, authors analysed the uncertainties induced by different definitions of the momentum 
fraction $z$, including the covariant and non-covariant definitions and showed that the FFs corresponding 
to the light-cone definition of  the fragmentation  parameter $z$ are equivalent to the ones
in the infinite momentum frame of gluon. Instead, the non-covariant definition (\ref{parameter})
is used as an approximation  unless $|\vec{k}|\rightarrow \infty$.\\
With the large heavy quark mass, the relative motion
of the heavy quark pair inside the charmonium is effectively nonrelativistic \cite{Ma:2013yla}.
For example, the squared relative velocity of the heavy quark pair in the quarkonium rest frame is 
$v^2\approx 0.22$ for the $J/\psi$ and $v^2\approx 0.1$ for the $\Upsilon$ \cite{Bodwin:2012xc}.
The nonrelativistic assumption allows us to use a simple mesonic wave function which is
the solution of the Schrodinger equation with a Coulomb potential \cite{Nejad:2013vsa}.
A typical simple mesonic wave function is given in \cite{brodsky} which is the 
nonrelativistic limit of the Bethe-Salpeter equation with the QCD kernel.
Here, according to the Lepage-Brodsky's approach \cite{Lepage:1980fj} we also neglect the relative motion of  the heavy
quark pair inside the charmonium states and we assume, for simplicity, that
the quark pair are emitted collinearly with each other and move along the $Z$-axes.
This assumption allows us to estimate the nonrelativistic mesonic wave function as a delta function 
form (more detail can be found in \cite{Nejad:2013vsa}).

\subsection{$J/\psi$ In gluon fragmentation}

\begin{figure}
\begin{center}
\includegraphics[width=0.5\linewidth,bb=260 535 420 700]{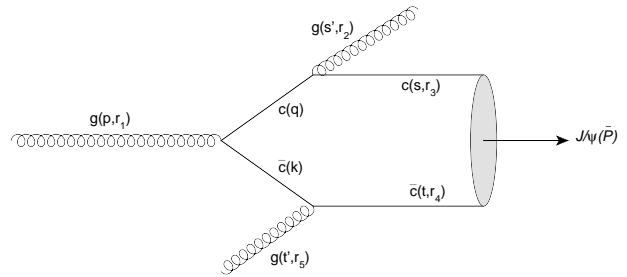}
\caption{\label{feyn2}%
Feynman diagram of process $g \rightarrow J/\psi gg$.}
\end{center}
\end{figure}
To derive an analytical result of the function $D_g^{J/\psi}(z, \mu)$ which refers
to the fragmentation of a gluon into the $J/\psi$,
we consider the Feynman diagram shown in Fig.~\ref{feyn2}, which is the subprocess
of the physical process presented in Fig.~\ref{feynman}. 
The spins ($r_i$) and the four-momenta of meson and partons are also labelled in Fig.~\ref{feyn2}.
According to our early assumption, the meson and its constituent quarks move along the Z-axes (fragmentation axes).
Then we set the relevant four-momenta as
\begin{eqnarray}\label{kinema}
s_\mu =[s_0, \bold{0}, s_L]                                &&\quad      t_\mu=[t_0, \bold{0}, t_L] \nonumber\\
s_\mu^\prime =[s_0^\prime, \bold{s_T^\prime}, s_L^\prime]  &&\quad     t_\mu^\prime=[t_0^\prime, \bold{t_T^\prime}, t_L^\prime]\nonumber\\
p_\mu=[p_0, \bold{p_T}, p_L]                              &&\quad     \bar P_\mu=[\bar P_0, \bold{0}, \bar {P}_L].
\end{eqnarray}
Considering the definition of fragmentation parameter, $z=E^H/E^g=\bar P_0/p_0$ (\ref{parameter}), we
also may write the parton energies in terms of the  initial gluon energy $p_0$  as
\begin{eqnarray}\label{kinematics}
s_0=x_1 z p_0  &&\quad  t_0=x_2 z p_0 \nonumber\\
s_0^\prime=x_3 (1-z) p_0   &&\quad   t_0^\prime=x_4 (1-z) p_0,
\end{eqnarray}
where $x_1=s_0/\bar{P}_0$ and $x_2=t_0/\bar{P}_0$ are  the meson energy fractions carried by the constituent quarks
which the condition $x_1+x_2=1$ holds for them.
Following Refs.~\cite{Kolodziej:1994uu,GomshiNobary:1994eq}, 
in the nonrelativistic approximation we assume  that the contribution of each 
constituent quark from the meson energy is proportional
to its mass, i.e. $x_1=m_c/M$ and $x_2=m_{\bar c}/M$ where $M=m_{J/\psi}$.
Furthermore, we assume that the two gluon jets move almost in the same direction.
This assumption is justified due to the fact that the very high momentum of the initial gluon
is predominantly carried in the forward direction.
Due to momentum conservation, the total
transverse momentum of the two gluon jets will be identical
to the transverse momentum of the initial gluon.
Therefore, we have $\bold{t_T^\prime}=\bold{s_T^\prime}=\bold{p_T}/2$.
By this assumption one also has $x_3=x_4=1/2$ in (\ref{kinematics}).\\
In the Suzuki's model the fragmentation function $D_g^{J/\psi}(z, \mu)$ at the initial
scale $\mu_0=m_{J/\psi}$, is obtained by squaring the total amplitude and integrating 
over final state phase space \cite{Suzuki:1985up,GomshiNobary:1994eq}
\begin{eqnarray}\label{first}
D_g^{J/\psi}(z, \mu_0)&=&\frac{1}{1+2r_1}\sum_s\int |T_M|^2\nonumber\\
&& \times\delta^3(\bold{\bar P}+\bold{s^\prime}+\bold{t^\prime}-\bold{p})d^3\bold{\bar P}d^3\bold{s^\prime}d^3\bold{t^\prime},
\end{eqnarray}
where, $r_1$ is the spin of initial gluon and $T_M$ is the probability amplitude of the $J/\psi$ production.
The amplitude $T_M$
involves the hard scattering amplitude $T_H$, which can be computed perturbatively from partonic
subprocesses, and the process-independent distribution amplitude $\Phi_M$ which contains 
the bound state nonperturbative dynamic of outgoing meson, i.e.
\begin{eqnarray}\label{base}
T_M(\bold{\bar P}, p, s^\prime, t^\prime)=\int [dx_i] T_H(\bold{\bar P}, p, s^\prime, t^\prime, x_i) \Phi_M(x_i, Q^2),\nonumber\\
\end{eqnarray}
where, $[dx_i]=dx_1dx_2\delta(1-x_1-x_2)$ and $x_i$'s are the momentum fractions carried by the constituent quarks.
This scheme,  introduced in \cite{Adamov:1997yk,brodsky}, is applied to absorb the soft behaviour of the 
bound state into the scattering amplitude $T_H$.
The amplitude $T_H$ is, in essence, the partonic cross section to produce a heavy quark
pair $c\bar c$ with certain quantum numbers that, in the old fashioned perturbation theory is expressed as
\begin{eqnarray}\label{first2}
T_H=\frac{(4\pi\alpha_s(2m_c))^{\frac{3}{2}} m_c^2 C_F}{2\sqrt{2\bar{P}_0 p_0 s_0^\prime t_0^\prime}}
\frac{\Gamma}{(\bar{P}_0+t_0^\prime+s_0^\prime-p_0)},
\end{eqnarray}
where  $\alpha_s$ is  the strong coupling constant, $C_F=\sqrt{5}/12$ is the color factor
and $\Gamma$ represents an appropriate combination of the quark propagators and the spinorial 
parts of the amplitude. It reads
\begin{eqnarray}\label{second}
\Gamma=G_1G_2\bigg\{\bar{u}(s, r_3) \displaystyle{\not}\epsilon_2^\star 
 (\displaystyle{\not}{q}+m_c)\displaystyle{\not}\epsilon_1 (\displaystyle{\not}{k}+m_c)\displaystyle{\not}\epsilon_5^\star v(t, r_4)\bigg\}.
\end{eqnarray}
Here, $\epsilon_i$'s are the gluon polarization vectors, $G_1=1/(q^2-m_c^2)=1/(2s.s^\prime)$ and
$G_2=1/(k^2-m_c^2)=1/(2t.t^\prime)$ are proportional to the quark propagators.
We put the dot products of the relevant four-vectors in the following form
\begin{eqnarray}
t.t^\prime=s.s^\prime=\frac{zm_c}{4M(1-z)}p_T^2+\frac{M(1-z)m_c}{4z}.
\end{eqnarray} 
In (\ref{base}),  $\Phi_M$ is the process-independent probability amplitude 
to find quarks co-linear up to a scale $Q^2$ in the mesonic bound state, 
so that by working in the infinite-momentum frame it can be estimated as a delta function.
Therefore, the distribution amplitude  for a S-wave heavy meson with neglecting the Fermi motion, reads \cite{brodsky,Amiri:1985mm}
\begin{eqnarray}\label{wave}
\Phi_M\approx\frac{f_M}{2\sqrt{3}} \delta(x_1-\frac{m_c}{M}),
\end{eqnarray}
where $M$ is the meson mass and $f_M=\sqrt{12/M}|\Psi(0)|$ is the meson decay constant which is related to the 
nonrelativistic mesonic S-wave function $\Psi(0)$ at the origin.
Substituting Eqs.~(\ref{base}), (\ref{first2}) and (\ref{wave}) in (\ref{first}) and carrying out the necessary
integrations, the fragmentation function  reads
\begin{eqnarray}\label{int}
D_g^{J/\psi}(z, \mu_0)&=&\frac{2}{3}(\pi\alpha_s)^3 (f_M C_F m_c^2)^2\nonumber\\
&&\hspace{-1cm}\times\int\frac{\frac{1}{3}\sum_s \bar{\Gamma}\Gamma
\delta^3(\bold{\bar{P}}+\bold{s}^\prime+\bold{t}^\prime-\bold{p})}{(\bar{P}_0 p_0 s_0^\prime t_0^\prime)
(\bar{P}_0+s_0^\prime+t_0^\prime-p_0)^2}
d^3\bold{\bar{P}} d^3\bold{s}^\prime d^3\bold{t}^\prime.\nonumber\\
\end{eqnarray}
To obtain an  analytical form of the fragmentation function for gluon to split into
the S-wave charmonia (i.e. $\eta_c, J/\psi$), we apply the scenario introduced in \cite{Guberina:1980dc}.
According to this scenario if $v(t)$ and $\bar{u}(s)$ are the Dirac spinors of the quarks forming the 
charmonium bound states,  in the nonrelativistic approximation the 
projection operator is defined as
\begin{eqnarray}\label{spin2}
\Lambda_{S,S_z}(s, t)=v(t)\bar{u}(s)\propto (\displaystyle{\not}{t}+m_c)\Pi_{S, S_z},
\end{eqnarray} 
where   $\Pi_{S S_z}$ is the appropriate spin projection operator; $\Pi_{00}=\gamma_5$
for pseudoscalar state ($\eta_c$) and $\Pi_{1S_z}=\displaystyle{\not}{\epsilon}(S_z)$ for vector state ($J/\psi$).
The spin content of the polarized meson is then given by either $\gamma_5$ or $\displaystyle{\not}{\epsilon}$, as could well be expected.
This operator is convenient for our assumption in which we ignore the Fermi motion, 
so that the constituent quarks  will fly together in  parallel.
Therefore, the spinorial part of the amplitude for formation of the vector charmonium state $J/\psi$, 
which we denote by $V$, may be presented in the following form
\begin{eqnarray}\label{firstscenarion}
\Gamma^V\propto G_1G_2\bigg\{(\displaystyle{\not}{t}+m_c)\displaystyle{\not}\epsilon\displaystyle{\not}\epsilon_2^\star 
(\displaystyle{\not}{q}+m_c)\displaystyle{\not}\epsilon_1(\displaystyle{\not}{k}+m_c)\displaystyle{\not}\epsilon_5^\star\bigg\},
\end{eqnarray}
where $q=s+s^\prime$ and $k=t+t^\prime$ are the energy-momenta of the virtual intermediate quarks, 
$\epsilon$ is the polarization four-vector of the meson $J/\psi$ which may be in a longitudinal state
$\epsilon^{(L)\mu}=\epsilon^\mu(\bar{P}, \lambda=0 )$ or a transverse 
state $\epsilon^{(T)\mu}=\epsilon^\mu(\bar{P}, \lambda=\pm 1)$.
These components satisfy the relations; $\epsilon(\bar{P}, \lambda).\bar{P}=0\quad$, 
$\bold\epsilon^{(T)}.\bold{\bar{P}}=0=\bold\epsilon^{(L)}\times\bold{\bar{P}}$ and 
$\bold\epsilon(\bar{P}, \lambda).\bold\epsilon^\star(\bar{P}, \lambda^\prime)=-\delta_{\lambda, \lambda^\prime}$.
Therefore, for a vector charmonium with the four-momentum $\bar P^\mu=[\bar P_0; \bold{0}, \bar P_L]$, the polarization four-vector 
is expressed as 
\begin{eqnarray}
\epsilon^{(L)\mu}&=&\frac{1}{M}(\bar{P}_L; 0, 0, \bar{P}_0),\nonumber\\
\epsilon^{(T)\mu}&=&\frac{1}{\sqrt{2}}(0; \mp 1, -i, 0),
\end{eqnarray}
where $\bar{P}_L=s_L+t_L$ and $\bar{P}_0=s_0+t_0$.\\
Now to obtain an  analytical form of the $J/\psi$ FF, in (\ref{int})
we perform a sum over the colors and the spins of  gluons. Then
the amplitude squared $\Gamma\bar \Gamma$  reads
\begin{eqnarray}\label{fafi}
\sum_{s}\Gamma^{V}\bar{\Gamma}^{V}&=&4G_1^2G_2^2Tr\big[(\displaystyle{\not}{t}+m_c)\displaystyle{\not}\epsilon
\gamma^\mu(\displaystyle{\not}{q}+m_c)\gamma^\nu(\displaystyle{\not}{k}+m_c)\nonumber\\
&&\times (\displaystyle{\not}{k}+m_c)\gamma_{\nu}(\displaystyle{\not}{q}+m_c)
\gamma_{\mu}\displaystyle{\not}\epsilon^\star(\displaystyle{\not}{t}+m_c)\big].
\end{eqnarray}
Using the traditional trace technique, the trace may be expressed as the dot products of four-vectors.
Here, we put the dot products of the relevant four-vectors in the following forms:
\begin{eqnarray}
t^\prime.s^\prime&=&0,\nonumber\\
p.\epsilon_T&=&-\frac{p_T}{\sqrt{2}}(-1+i),\nonumber\\
p.\epsilon_L&=&-\frac{M}{2z}+\frac{z}{2M}p_T^2,\nonumber\\
p.s&=&p.t=\frac{m_c^2}{z}+\frac{z}{4}p_T^2,\nonumber\\
p.t^\prime&=&p.s^\prime=\frac{z^2}{4(1-z)}p_T^2,\nonumber\\
s^\prime.\epsilon_T&=&\epsilon_T.t^\prime=-\frac{p_T}{2\sqrt{2}}(-1-i),\nonumber\\
t.s^\prime&=&s.t^\prime=\frac{1-z}{2z}m_c^2+\frac{z}{8(1-z)}p_T^2,\nonumber\\
s^\prime.\epsilon_L&=&\epsilon_L.t^\prime=-\frac{M(1-z)}{4z}+\frac{z}{4M(1-z)}p_T^2.
\end{eqnarray}
To proceed we need to specify the phase space integrations in (\ref{int}). Note that
\begin{eqnarray}
&&\int \frac{d^3\bold{\bar{P}}\delta^3(\bold{\bar{P}}+\bold{t^\prime}+
\bold{s^\prime}-\bold{p})}{\bar{P}_0 (\bar{P}_0+s_0^\prime+t_0^\prime-p_0)^2}=\nonumber\\
&&\hspace{+1cm}\frac{\bar{P}_0}{(M^2+2p.t^\prime+2p.s^\prime-2s^\prime.t^\prime)^2}.
\end{eqnarray}
Here, instead of performing the transverse momentum integrations we replace the integration variable
by its average value $ \left\langle p_T^2 \right\rangle$ in each case, 
which is a free parameter and can be specified experimentally. Therefore we can write
\begin{eqnarray}
\int{ F(z, \bold{s_T^{\prime}}) d^3 s^\prime}&=&\int{F(z, \bold{s_T^{\prime}}) ds_L^\prime d^2s_T^\prime} \nonumber\\
&\approx & m_c^2 s_0^\prime F(z, \left\langle s_T^{\prime 2}\right\rangle)=
m_c^2 s_0^\prime F(z, \frac{1}{4}\left\langle p_T^2\right\rangle),\nonumber\\
\end{eqnarray}
and
\begin{eqnarray}
\int{ F^\prime(z, \bold{t_T^{\prime}}) d^3 t^{\prime}}&=&\int{F^\prime(z, \bold{t_T^{\prime}}) dt_L^\prime d^2t_T^\prime} \nonumber\\
&\approx & m_c^2 t_0^\prime F^\prime(z, \left\langle t_T^{\prime 2}\right\rangle)=
m_c^2 t_0^\prime F^\prime(z, \frac{1}{4}\left\langle p_T^2\right\rangle).\nonumber\\
\end{eqnarray}
Finally, putting all in (\ref{int}) and by assuming $M=2m_c$ in the nonrelativistic limit, we obtain 
the longitudinal and the transverse components of the $g\rightarrow J/\psi$ FF, as
\begin{eqnarray}\label{d1}
D_{g\rightarrow J/\psi}^{T}(z, \mu_0)&=&\frac{8 N_1 z \alpha_s^3}{g(z, \left\langle p_T^{2}\right\rangle)} 
\bigg[-16\frac{1-z^2}{z^2}m_c^6\nonumber\\
&&-\frac{8}{z(1-z)}\big(2z^3+4z^2+3z-4\big)m_c^4p_T^2\nonumber\\
&&\hspace{-1cm}+\frac{1}{(1-z)^3}\bigg(z(17z^2-25 z+8)m_c^2p_T^4-z^4p_T^6\bigg)\bigg],\nonumber\\
\end{eqnarray}
and 
\begin{eqnarray}\label{d2}
D_{g\rightarrow J/\psi}^{L}(z, \mu_0)&=&\frac{4 N_1 z\alpha_s^3}{g(z, \left\langle p_T^{2}\right\rangle)} 
\bigg[-32\frac{1-z^2}{z^2}m_c^6\nonumber\\
&&+\frac{16}{1-z}\big(3z^2+2z+2\big)m_c^4p_T^2+\nonumber\\
&&\hspace{-1.3cm}\frac{1}{(1-z)^3}\bigg(2z^2(-4z^2-7z+11)m_c^2p_T^4+3z^4p_T^6\bigg)\bigg].\nonumber\\
\end{eqnarray}
where $N_1=(\pi^3 m_c^8 f_M^4 C_F^2 )/216$, and
\begin{eqnarray}
g(z, \left\langle p_T^{2}\right\rangle)&=&\bigg[\frac{zm_c}{2M(1-z)}p_T^2+\frac{M(1-z)m_c}{2z}\bigg]^4\nonumber\\
&&\times \bigg[M^2+\frac{z^2}{1-z}p_T^2\bigg]^2.
\end{eqnarray}
Note that, the fragmentation function for a vector charmonium $J/\psi$ is the sum of the longitudinal and twice the
transverse components, i.e. $D_{g}^{J/\psi}(z, \mu_0)=2D_{g}^{T}+D_{g}^{L}$.\\
Finally, we point out that to impose the effect of quarkonium spin into the calculation, 
a second scenario is defined in \cite{Kolodziej:1994uu,Kuhn:1979bb}, where 
 in the nonrelativistic approximation the spin projection operator is expressed as
\begin{eqnarray}\label{spin1}
\Lambda_{S,S_z}(\bar{P})=\frac{f_M}{\sqrt{48}}(\displaystyle{\not}{\bar{P}}+M)\Pi_{SS_z},
\end{eqnarray} 
where, $f_M$ is the meson decay constant
and $M$ and $\bar{P}$ are the mass and the four-momentum of  meson bound state, respectively.
In the nonrelativistic limit $(\displaystyle{\not}{\bar{P}}=2\displaystyle{\not}{t}, M=2 m_c)$, these two scenarios (\ref{spin2}, \ref{spin1}) 
are the same and, in conclusion, the sum of transverse and longitudinal polarisations of the $J/\psi$
should be identical.

\subsection{$\eta_c$ In gluon fragmentation}

\begin{figure}
\begin{center}
\includegraphics[width=0.8\linewidth,bb=190 535 420 700]{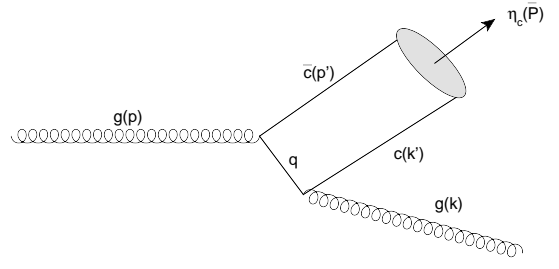}
\caption{\label{feyn1}%
Feynman diagram of process $g \rightarrow \eta_c$ at $\alpha_s^2$-order.}
\end{center}
\end{figure}
Following Ref.~\cite{Braaten:1993rw}, the fragmentation function for a gluon to split
into the pseudoscalar S-wave charmonium $\eta_c$ can be calculated by considering Fig.~\ref{feynmangraph}a.
To derive an analytical result of the $g\rightarrow \eta_c$ FF we consider the Feynman diagram 
shown in Fig.~\ref{feyn1}, where a gluon forms a bound state $c\bar c$ with a gluon produced through a single c-quark.
The four-momenta of meson and partons are also labelled.
According to our previous assumption the meson and its constituent quarks move along the Z-axes (fragmentation axes).
Then one can set the relevant four-momenta as
\begin{eqnarray}\label{kinemat}
p_\mu =[p_0, \bold{p_T}, p_L]                                &&\quad      k_\mu=[k_0, \bold{p_T}, k_L] \nonumber\\
p_\mu^\prime =[p_0^\prime, \bold{0}, p_L^\prime]  &&\quad     k_\mu^\prime=[k_0^\prime, \bold{0}, k_L^\prime]\nonumber\\
&&\hspace{-1cm}    \bar P_\mu=[\bar P_0, \bold{0}, \bar {P_L}].
\end{eqnarray}
Considering the definition of fragmentation parameter (\ref{parameter}), we
may write the parton energies in terms of the gluon energy $p_0$  as; 
$p_0^\prime=x_1 z p_0, k_0^\prime=x_2 z p_0, k_0=(1-z)p_0$
where $x_1$ and $x_2$ are  the meson energy fractions carried by the constituent quarks.
Following our early discussion, the contribution of each 
constituent quark from the meson energy is proportional
to its mass, i.e. $x_1=m_c/M$ and $x_2=m_{\bar c}/M$ where $M=m_{J/\psi}$.\\
We start with the definition of the fragmentation function as
\begin{eqnarray}\label{basic}
D_g^{\eta_c}=\frac{1}{1+2s_g}\sum_s\int |T_M|^2\delta^3(\bold{\bar P}+\bold{k}-\bold{p})d^3\bold{\bar P}d^3\bold{k},
\end{eqnarray}
where the probability amplitude $T_M$ is related to the hard scattering amplitude $T_H$
and the distribution amplitude $\Phi_M$ as in (\ref{base}).
In view of our early discussion in this section, we propose a delta function as (\ref{wave})
for the amplitude $\Phi_M$.\\
Using the perturbation theory, the amplitude $T_H$ is written in the following form
\begin{eqnarray}\label{sec1}
T_H=\frac{4\pi\alpha_s(2m_c) m_c^2 C_F}{2\sqrt{2\bar{P}_0 p_0 k_0}}
\frac{\Gamma}{(\bar{P}_0+k_0-p_0)},
\end{eqnarray}
where  $\Gamma$ represents the spinorial parts of the amplitude as
\begin{eqnarray}\label{sec2}
\Gamma=G\bigg\{\bar{u}(k^\prime, r_2) \displaystyle{\not}\epsilon_2^\star(k) 
 (\displaystyle{\not}{q}+m_c)\displaystyle{\not}\epsilon_1(p) v(p^\prime, r_1)\bigg\}.
\end{eqnarray}
Here, $G=1/(q^2-m_c^2)=1/(2k.k^\prime)$ is proportional to the quark propagator and
$C_F=1/(2\sqrt{3})$ is the color factor.

Putting all in (\ref{basic}), the fragmentation function reads
\begin{eqnarray}\label{inttt}
D_g^{\eta_c}(z, \mu_0)&=&\frac{1}{6}(\pi\alpha_s f_M C_F m_c^2)^2\nonumber\\
&&\hspace{-1cm}\times\int \frac{\frac{1}{2}\sum_s \bar{\Gamma}\Gamma}{p_0 k_0}d^3\bold{k}
\int\frac{\delta^3(\bold{\bar{P}}+\bold{k}-\bold{p})}{\bar{P}_0(\bar{P}_0+k_0-p_0)^2}d^3\bold{\bar{P}}.\nonumber\\
\end{eqnarray}
Considering the scenario introduced in previous section (\ref{spin2}), where 
the spin projection operator for a pseudoscalar charmonium state $\eta_c$ is
$\Pi_{00}=\gamma_5$, the spinorial part of the amplitude (\ref{sec2}) 
is presented in the following form
\begin{eqnarray}
\Gamma^P\propto G\bigg\{(\displaystyle{\not}{k^\prime}+m_c)\gamma_5\displaystyle{\not}\epsilon_2^\star(k) 
 (\displaystyle{\not}{q}+m_c)\displaystyle{\not}\epsilon_1(p)\bigg\}.
\end{eqnarray}
By performing a sum over the colors and the spins of gluons, the amplitude squared reads
\begin{eqnarray}
\sum_{s}\Gamma^{P}\bar{\Gamma}^{P}=128G^2m_c^2\bigg\{2m_c^2+2k.k^\prime+p^\prime.(k+k^\prime)\bigg\}.
\end{eqnarray}
Next we consider the phase space integrations (\ref{inttt}). 
Following our previous approach, we have
\begin{eqnarray}
\int \frac{d^3\bold{\bar{P}}\delta^3(\bold{\bar{P}}+
\bold{k}-\bold{p})}{\bar{P}_0 (\bar{P}_0+k_0-p_0)^2}
=\frac{\bar{P}_0}{(M^2+2p.k)^2},
\end{eqnarray}
and 
\begin{eqnarray}
\int{ F(z, p_T^2) d^3 \bold{k}}= m_c^2 k_0 F(z, \left\langle p_T^{2}\right\rangle).
\end{eqnarray}
In order to get the correct result for the fragmentation function of the process $g\rightarrow \eta_c$,
it is necessary to consider a second diagram that can be obtained from Fig.~\ref{feyn1} by 
interchanging the two vertices where the gluons attached to the heavy quark lines.\\
In conclusion, in the referred scenario the fragmentation function of the process $g\rightarrow \eta_c$ 
is expressed as 
\begin{eqnarray}\label{b2}
D_{g\rightarrow \eta_c}(z, \mu_0)&=&\frac{N_2 \alpha_s^2 z}{g(z, \left\langle p_T^{2}\right\rangle)}(192 m_c)\nonumber\\
&&\times\bigg[2m_c+\frac{z p_T^2}{M(1-z)}+\frac{M(1-z)}{z}\bigg].\nonumber\\
\end{eqnarray}
where $N_2=(\pi m_c^3f_M^2C_F)^2/864$, and
\begin{eqnarray}
g(z, \left\langle p_T^{2}\right\rangle)=\bigg[(M^2+\frac{z^2 p_T^2}{1-z})
(\frac{z^2 p_T^{2}+M^2(1-z)^2}{Mz(1-z)})\bigg]^2.\nonumber\\
\end{eqnarray}

\section{Results and discussion}
\label{sec:three}

\begin{figure}
\begin{center}
\includegraphics[width=1\linewidth,bb=88 570 322 730]{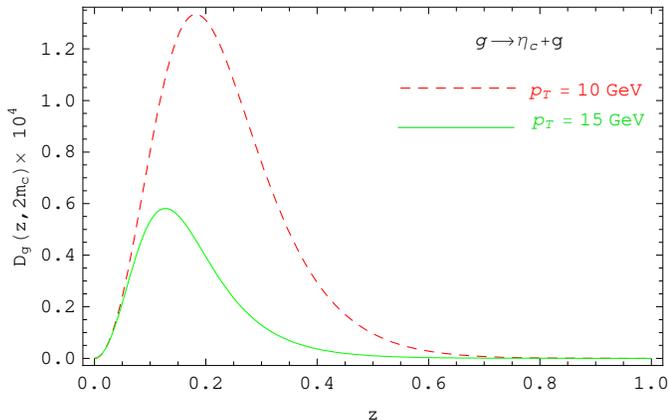}
\caption{\label{fig2}%
The fragmentation function for the process $g\rightarrow \eta_c$, considering two
values of the transverse momentum $p_T=10, 15$ GeV.
The initial scale is set as $\mu_0=2m_c$.}
\end{center}
\end{figure}
\begin{figure}
\begin{center}
\includegraphics[width=1\linewidth,bb=88 570 322 730]{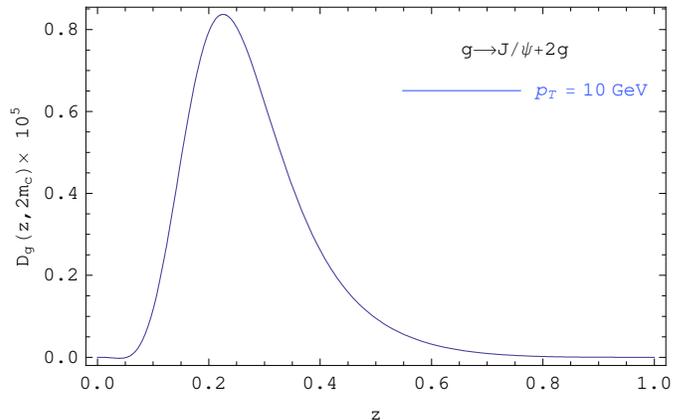}
\caption{\label{fig3}%
The fragmentation function for the process $g\rightarrow J/\psi$ at $\alpha_s^3$.
Here, $D_{g}^{J/\psi}(z, \mu_0)=2D_{g}^{T}+D_{g}^{L}$ where the transverse component
$D_{g}^{T}$ is given in (\ref{d1}) and the longitudinal one $D_{g}^{L}$ is given in (\ref{d2}).}
\end{center}
\end{figure}

We are now in a position to present our phenomenological predictions for 
the gluon fragmentation into the $\eta_c$ and $J/\psi$, by performing a numerical analysis.
In general, the fragmentation function $D_{g}^H(z,\mu)$ depends on the factorization scale $\mu$.
Here, we have set the scale in the FF and
in the running coupling constant $\alpha_s(\mu)$ to $\mu=2m_c$ which is
the minimum value of the invariant mass of the fragmenting gluon. 
Therefore, the functions (\ref{d1}, \ref{d2}) and (\ref{b2}) should be regarded as models for 
the $g\rightarrow \eta_c, J/\psi$ FFs at the initial scale $\mu_0=2m_c$.
For values of $\mu$ much larger than $\mu_0$, the initial FFs should be evolved from the
scale $\mu_0$ to the higher scale $\mu$ using the Altarelli-Parisi equations \cite{dglap1, dglap2}
\begin{eqnarray}
\mu\frac{\partial}{\partial \mu}D_{i\rightarrow H}(z, \mu)=\sum_j\int_z^1\frac{dy}{y}
P_{i\rightarrow j}(z/y, \mu)D_{j\rightarrow H}(y, \mu),\nonumber\\
\end{eqnarray}
where $P_{i\rightarrow j}(x, \mu)$ are the Altarelli-Parisi functions for the
splitting of the parton of type $i$ into a parton of type $j$
with momentum fraction $x$. The only boundary condition on this evolution equation is the fragmentation
function $D_{i\rightarrow H}(z, \mu_0)$ at the scale $\mu_0=2m_c$.\\
For numerical results, we take $m_c=1.5$ GeV, $m_{J/\psi}=3096.9$ MeV, $m_{\eta_c}=2983.6$ MeV, 
 $\alpha_s(2m_c)=0.26$ and $f_M(c\bar{c})=0.48$ GeV \cite{Gomshi}.\\
In Fig.~\ref{fig2},  our prediction for 
the $g\rightarrow \eta_c$ FF (\ref{b2}) by considering  two values of the gluon transverse momentum 
($p_T=10, 15$ GeV) is shown.
It shows that the fragmentation function distribution relies on the momentum $p_T$ of the initial gluon.
In Fig.~\ref{fig3}, the behaviour of the $g\rightarrow J/\psi$ FF is studied.
Here, $D_g^{J/\psi}$ is the convenient summation of the longitudinal and 
transverse fragmentation functions as $D_{g}^{J/\psi}(z, \mu_0)=2D_{g}^{T}+D_{g}^{L}$, 
see Eqs.~(\ref{d1}, \ref{d2}). 
Our result shown in Fig.~\ref{fig3}, is in acceptable agreement  with the result presented
in Fig.~3 of Ref.~\cite{Qi:2007sf}, when the non-covariant definition
of fragmentation parameter (\ref{parameter}) is applied.
 In both results, the peak position of the FF occurs at $z\approx 0.22$ 
when $p_T=10$ GeV is considered, and
the maximum value of the FF is $D_g^{J/\psi}\approx 0.8\times 10^{-5}$.\\
Besides the $g\rightarrow J/\psi$ FF itself, also its first two moments are of phenomenological interest.
They correspond to the $g\rightarrow J/\psi$ branching fraction
\begin{eqnarray}
B(\mu)=\int_0^1 dz D_{g\rightarrow J/\psi}(z, \mu),
\end{eqnarray}
and the average energy fraction that the $J/\psi$ meson receives from the gluon
\begin{eqnarray}
\left\langle z\right\rangle(\mu)=\frac{1}{B(\mu)}\int_0^1 z dz D_{g\rightarrow J/\psi}(z, \mu).
\end{eqnarray}
Indeed, an order of magnitude estimate of the gluon fragmentation contribution to $J/\psi$ production in
any high transverse momentum process can be obtained by multiplying the cross section for producing gluons
with $p_T>2m_c$ by the branching fraction, which refers to the fragmentation probability.
Our result for the $g\rightarrow J/\psi$ branching fraction
is  $B_{J/\psi}(2m_c)=2.94\times 10^{-6}$ and
$\left\langle z\right\rangle_{J/\psi}(2m_c)=0.277$ which this initial fragmentation 
probability can be compared with the result presented in \cite{Braaten:1993rw} where $B(2m_c)=3.2\times 10^{-6}$.

Our results may be directly applied to the S-wave bottomonium sates $\Upsilon (^{3}S_1)$ and $\eta_b(^{1}S_0)$, except 
that $m_c$ is replaced by $m_b=4.5$ GeV and the decay constant $f_M(b\bar{b})=0.33$ GeV \cite{Gomshi}
is the appropriate constant for the bottomonium states.
Since the b-quark is heavier than the c-quark, we expect that the peaks of
the fragmentation functions shift significantly toward higher values of z.
Our results for the $g\rightarrow \eta_c, J/\psi$ FFs will be checked by the comparison 
between the theoretical predictions and the experimental measurements of the heavy meson cross sections at the LHC.

\section{Conclusion}
\label{sec:four}

Understanding hadronization, the process by which a parton 
evolves into a hadron, is complicated by its intrinsically nonperturbative nature.
In hadron colliders, at sufficiently large transverse momentum of the heavy quarkonium production
the direct production schemes are normally suppressed  while the fragmentation mechanism becomes dominant.
The fragmentation refers to the process  of a parton with high transverse momentum which subsequently decays
into the expected hadron \cite{Braaten:1993rw}.  Beside the phenomenological approaches,
it is well-known that the fragmentation function which  describes this process
can be calculated using perturbative QCD.
In this paper we, for the first time, gave out an analytical form for the leading color-singlet 
contribution to the fragmentation function for a gluon to split into the vector and pseudoscaler
S-wave charmonia ($J/\psi, \eta_c$) at the initial scale $\mu=2m_c$. 
We used a different model in getting them from the Braaten's model applied in 
other literatures, see Refs.~\cite{Braaten:1993rw,Qi:2007sf}.
Our results depend on the transverse momentum of the initial gluon and shows a simple analytical form whereas 
in the Braaten's model, the integrations over all freedom degrees are performed
and due to the lengthy and cumbersome expression of the FFs the results are presented 
as the two-dimensional integrals that must be evaluated numerically.
Since the transverse momentum dependent FFs show up explicitly in semi-inclusive cross sections, therefore
in the QCD corrections the inclusion of these dependent FFs will be necessary.
Our result for the $g\rightarrow J/\psi$ FF is in acceptable consistency  with the numerical result presented in Ref.~\cite{Qi:2007sf},
when one uses the normal definition of the fragmentation parameter (\ref{parameter}). 
We also found that the fragmentation
probability of a high energy gluon splitting into $J/\psi$ is about $10^{-6}$.

\end{document}